# The Peregrine breather on the zero-background limit as the two-soliton degenerate solution: An experimental study

**Amin Chabchoub[1,2*], Alexey Slunyaev[3,4], Norbert Hoffmann[5,6], Frédéric Dias[7,8], Bertrand Kibler[9], Goëry Genty[10], John M. Dudley[11] and Nail Akhmediev[12]**

[1]Centre for Wind, Waves and Water, School of Civil Engineering, The University of Sydney, Sydney, NSW 2006, Australia

[2]Marine Studies Institute, The University of Sydney, Sydney, NSW 2006, Australia

[3]Department of Nonlinear Geophysical Processes, Institute of Applied Physics RAS, Nizhny Novgorod, Russia

[4]Laboratory of Dynamical Systems and Applications, National Research University Higher School of Economics, Nizhny Novgorod, Russia

[5]Dynamics Group, Hamburg University of Technology, 21073 Hamburg, Germany

[6]Department of Mechanical Engineering Imperial College London, London SW7 2AZ, United Kingdom

[7]School of Mathematics and Statistics, University College Dublin, Dublin, Ireland

[8]Université Paris-Saclay, ENS Paris-Saclay, CNRS, Centre Borelli, 91190 Gif-sur-Yvette, France

[9]Laboratoire Interdisciplinaire Carnot de Bourgogne (ICB), UMR 6303 CNRS-Université Bourgogne Franche-Comté, 21078 Dijon, France

[10]Photonics Laboratory, Physics Unit, Tampere University, 33014 Tampere, Finland

[11]Institut FEMTO-ST, Université Bourgogne Franche-Comté CNRS UMR 6174, Besançon, France

[12]Department of Theoretical Physics, Research School of Physics, The Australian National University, Canberra, ACT 2600, Australia

**\* Correspondence:**
amin.chabchoub@sydney.edu.au

**Abstract**

Solitons are coherent structures that describe the nonlinear evolution of wave localizations in hydrodynamics, optics, plasma and Bose-Einstein condensates. While the Peregrine breather is known to amplify a single localized perturbation of a carrier wave of finite amplitude by a factor of three, there is a counterpart solution on zero background known as the degenerate two-soliton which also leads to high amplitude maxima. In this study, we report several observations of such multi-soliton with doubly-localized peaks in a water wave flume. The data collected in this experiment confirm the distinctive attainment of wave amplification by a factor of two in good agreement with the dynamics of the nonlinear Schrödinger equation solution. Advanced numerical simulations solving the problem of nonlinear free water surface boundary conditions of an ideal fluid quantify the physical limitations of the degenerate two-soliton in hydrodynamics.

## 1  Introduction

The Peregrine breather (PB) [1] is a fundamental solution of the nonlinear Schrödinger equation (NLSE) localized both in space and time, yielding a three-fold amplification of the initial amplitude at the point of maximum localization. These unique characteristics have led the PB to be generally considered as a potential backbone model allowing to describe the emergence of extreme events in several physical systems [2,3]. Although the PB existence was originally predicted in the early eighties [1], it took about three decades to observe this particular wave envelope in a laboratory environment [4,5,6]. These initial studies have attracted significant attention and led to many follow-up studies related to PB dynamics and its peculiar physical properties [10-17]. The initial or boundary conditions leading to the PB excitation require to impose a small perturbation on top of a plane wave background. Recently, generic features of PB dynamics on a stationary dnoidal background have been presented [18] and in fact the regular background represents only one limiting case of the exact NLSE family of dnoidal solutions while the other limit is the envelope soliton on zero-background [19,20]. This allows a more general construction of Peregrine-type coherent structures on different type of stationary backgrounds, which can be also described by an exact solution [21].

In this paper, we experimentally investigate the PB dynamics in the zero background limit, which can be also associated with the degenerate case of two soliton interaction, resulting in an amplitude amplification factor of two at the point of maximum localization [22]. The laboratory experiments, conducted in different water wave flumes are in excellent agreement with the theory when the carrier steepness is moderate. Otherwise, deviations from the symmetric envelope shapes are inevitable due to the physical limitations of the NLSE approach to describe broadband processes in water waves. The numerical simulations based on the higher-order spectral method, which accurately solve the nonlinear water wave problem, quantify the limitations in the evolution of the hydrodynamic degenerate soliton on the water surface. We believe that our results will have a significant impact on the field of nonlinear dynamics and improve fundamental understanding of extreme wave formation in nonlinear media.

## 2  Higher-order solitons on zero background and degeneracy

The NLSE for surface gravity waves is the simplest nonlinear evolution equation that takes into account the interplay between dispersion and nonlinearity in the evolution of a narrowband wave field. Assuming unidirectional propagation of the wave field in infinite water depth, the wave envelope evolution equation reads [23]

$$i\left(\frac{\partial \psi}{\partial t} + c_g \frac{\partial \psi}{\partial x}\right) + \frac{\omega}{8k^2}\frac{\partial^2 \psi}{\partial x^2} + \frac{\omega k^2}{2}|\psi|^2\psi = 0, \qquad (1)$$

where $\psi(x,t)$ is the complex wave envelope, $x$ is the spatial coordinate along the wave propagation, and $t$ represents time. The parameters $\omega$ and $k$ are the carrier cyclic wave frequency and wavenumber, respectively. The latter are constrained by the gravitational acceleration $g$-dependent deep-water dispersion equation

$$\omega^2 = gk, \qquad (2)$$

and the envelope is assumed to propagate with the group velocity $c_g = \frac{\partial \omega}{\partial k} = \frac{\omega}{2k}$.

The NLSE is a partial differential equation that belongs to the family of integrable evolution equations [24]. Its exact solutions provide physically-relevant models for investigating the dynamics of nonlinear coherent wave envelopes in controlled laboratory environments. The fields of its application are hydrodynamics, optics and Bose-Einstein condensates. It is common to use the dimensionless form of Eq. (1) in particular when aiming for the derivation of exact solutions

$$i\frac{\partial \Psi}{\partial T} + \frac{\partial^2 \Psi}{\partial X^2} + 2|\Psi|^2\Psi = 0, \qquad (3)$$

which is obtained by introducing the following transformations

$$X = 2k(x - c_g t), \quad T = \frac{\omega}{2}t, \quad \Psi = \frac{k}{\sqrt{2}}\psi. \qquad (4)$$

One of the most-fundamental solutions of the NLSE is an isolated sech-shape nonlinear wave group on zero-background known as envelope soliton, which can be considered as a mode of a nonlinear system [25] which remains unchanged with propagation. At the same time, interactions and collisions between envelope solitons are elastic [26,27]. The number of solitons contained in a localized initial condition remains fixed during the follow up evolution. The zero-velocity soliton solution with an amplitude of one can be written as

$$\Psi_S(X, T) = \text{sech}(X)\exp(iT). \qquad (5)$$

The initial-value problem for the NLSE can be solved with the help of the inverse scattering technique (IST) [24,28,29]. More complex (higher-order) structures containing multiple solitons can be also constructed using the Darboux transformation [30] or other dressing method [31]. Each envelope soliton in these superpositions is unambiguously characterized by the pair of its two key parameters: the amplitude and the velocity. The NLSE solution describing the dynamics of two envelope solitons with fixed amplitudes 0.5 and 1.5, zero-velocities and located at the same position $X=0$, is known as the Satsuma-Yajima breather [32]

$$\Psi_{S_2}(X, T) = 4\frac{\cosh 3X + 3\cosh X \exp 8iT}{\cosh 4X + 4\cosh 2X + 3\cosh 8T}\exp(iT). \qquad (6)$$

This solution is periodic in $T$ and can be used for pulse nonlinear wave group compression. At $T = 0$, this solution takes the form of a soliton with twice the amplitude of a single soliton of the same width, i.e. $\Psi_{S_2}(X, 0) = 2\,\text{sech}(X) = 2\,\Psi_S(X, 0)$. However, this initial condition changes with propagation and evolves towards self-compression. Such solutions also play a key role in the formation of significant irreversible spectral broadening and the creation of supercontinua as a result of soliton fission [33,34]. Generally, when the parameters of the two envelope solitons become close, the distance between them increases and they repel each other, moving away towards infinity. Due to this fact, for more than two decades since the development of the IST, the two-soliton solution of the NLSE with exactly the same parameters has been considered as non-existent. Overcoming this controversy, the solution has been reported in [1,22]. Such solution is the degenerate two-soliton solution, as finding it requires considering the special limit when their amplitudes and velocities tend to the same limiting values. It is represented by a mixed semi-rational semi-hyperbolic function

$$\Psi_D(X, T) = 4\frac{X\sinh X - \cosh X - 2iT\cosh X}{\cosh 2X + 1 + 2X^2 + 8T^2}\exp(iT). \qquad (7)$$

More specifically, the solution (7) describes the interaction of two envelope solitons with unit amplitudes and with their center of mass located at $X=0$. The envelope $|\Psi_D|$ in (7) is symmetric with respect to the change of the sign of either $X$ or $T$. Note that the solution (7) may be generalized using the invariant transforms of the NLSE, i.e. arbitrary phase, scaling and Galilean transforms. In the reduced form (7), it does not contain any free parameters. The degenerate solution (7) describes two

"attracting" envelope solitons when $T < 0$. When $T = 0$, the two solitons are superimposed and form an extreme event with an amplitude at the point of collision twice that the amplitude of the isolated solitons. At large times $T \gg 1$, the solution (7) describes the two envelope solitons which slowly walk away from each other after the collision. Each of them can be approximated as a quasi-single-soliton solution. The opposite velocities of the two solitons reduce when $T \to \infty$.

What at first sight seems to be a mathematical artifact, has in fact a particular physical relevance. Indeed, the central part of the degenerate solution (7) can be considered as the PB on the zero-background limit. The comparison is relevant because the solution (7) is semi-rational while PB is a rational solution. Representing hyperbolic functions $\cosh X$ and $\cosh 2X$ in the central part of the solution as an expanded series in $X$ can reduce it to a rational approximation similar to the PB. On the other hand, the PB can be excited on top of exact dnoidal solutions, parameterized as $\Psi_{dn}(X,T) = \text{dn}(X,m)\exp(i[2-m^2]T);\ 0 \le m \le 1$ see [18,21]. One limiting case of this one-parameter family of steady dnoidal solutions is the regular background ($m = 0$) and the other limit is the envelope soliton ($m = 1$). This second limit leads to the formation of the degenerate soliton solution. The transformation is controlled by an additional free parameter – modulus of the dn function. The role of this parameter in the highly nontrivial process of degenerate soliton formation can be seen from Fig. 7 in [18]. This process admits several stages of PB transformation. A significantly simplified version of the process can be seen from Fig. 1 (see bottom panels from left to right). Here, the classical Peregrine solution on finite background is transformed to the degenerate solution on zero background with one intermediate step in the form of the PB on the DN-wave background (referring to the semicircle in the $\lambda$-plane in Fig. 2 of [18]).

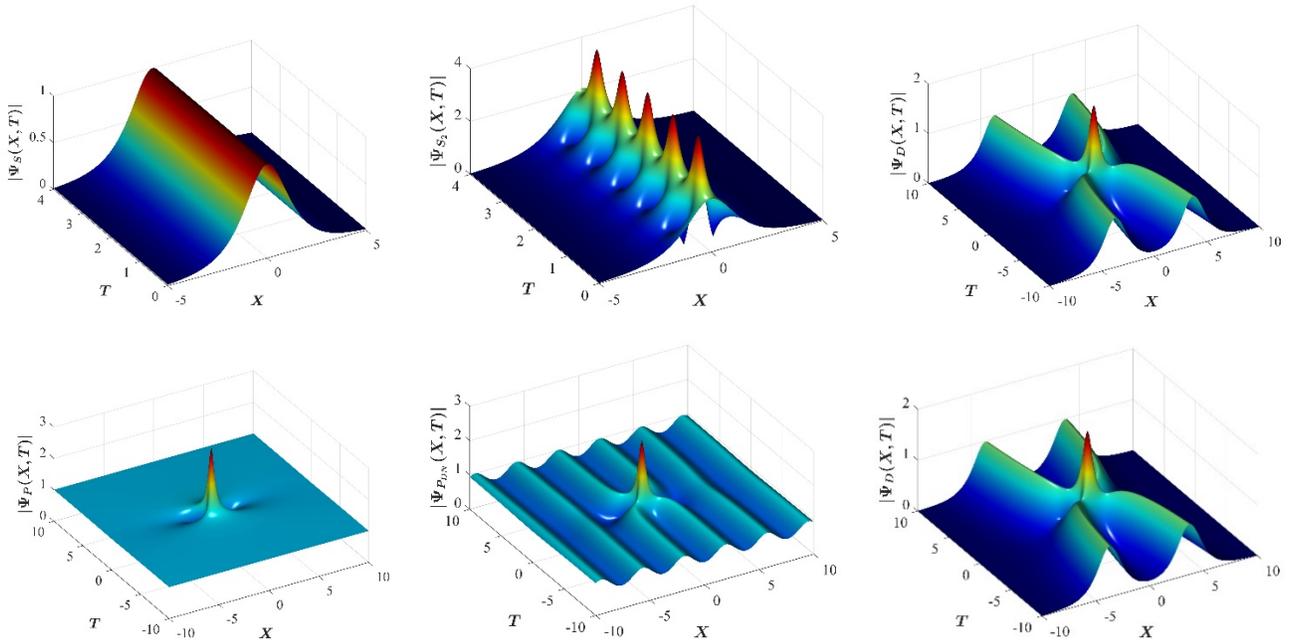

**Figure 1: Spatio-temporal evolution of solitons on finite and zero-background.** Top Left: single envelope soliton. Top Middle: Higher-order soliton of order 2. Top Right: Degenerate two-soliton solution. Bottom left: Peregrine breather. Bottom middle: Peregrine breather on a dnoidal background. Bottom right: Degenerate two-soliton soliton.

The Peregrine solution

$$\Psi_P(x,t) = \left(-1 + \frac{4+16iT}{1+4X^2+16T^2}\right)\exp(2iT), \tag{8}$$

has been found to be present in multi-soliton solutions [35]. On the other hand, the degenerate two-soliton solution (7) can represent Peregrine-type dynamics with zero condensate. This becomes more evident when considering the type of localization around point of maximum amplitude. In fact, the shape of the extreme wave at $T = 0$ does resemble the shape of the Peregrine breather. This can be seen from Fig. 3, where the degenerate two-soliton solution at $T = 0$ is compared with the shape of the Peregrine breather at $T = 0$ multiplied by the factor 2/3 in order to equalize the maximal amplitudes.

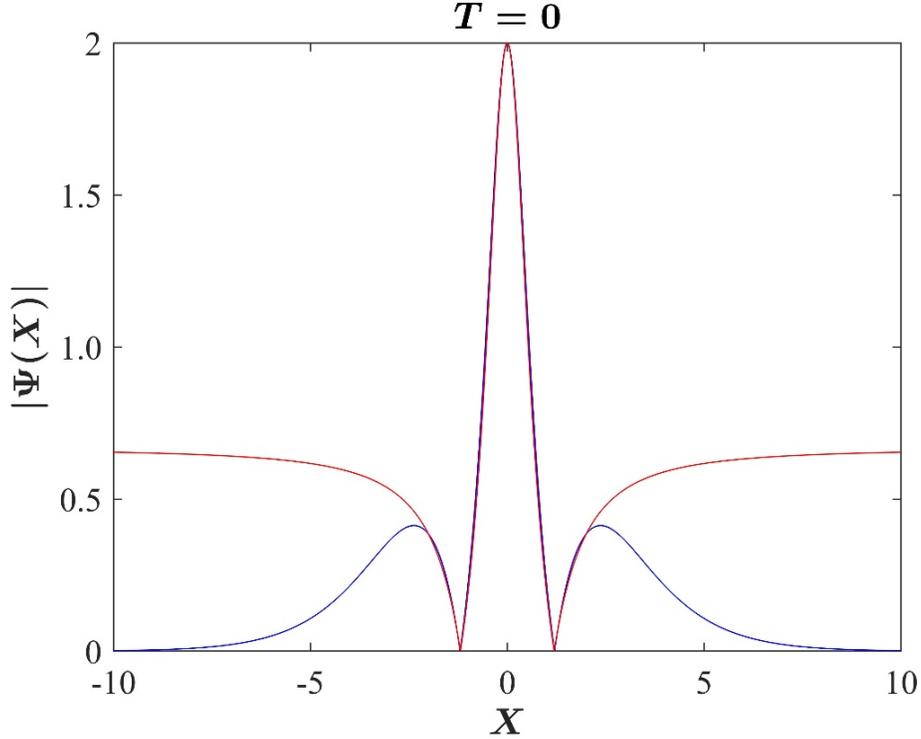

Figure 3: The amplitude rescaled Peregrine breather shape vs. degenerate two-soliton solution profile at *T=0*.

The agreement between the two profiles is remarkably good within the interval between the zeros.

Even though the dynamics of the degenerate two-soliton solution creates a smaller wave amplification than the Peregrine breather (2 rather than 3), it is still a rapidly forming extreme event. We should also take into account the difference between the backgrounds. Thus, such solutions can be responsible for the occurrence of extreme wave events, which are very similar to the PB.

## 3  Laboratory experiments

The physical experiments have been conducted in two different water wave facilities: Hamburg University of Technology and the University of Sydney flumes, as described in [10] and [36], respectively. Although both facilities are different when considering their size and type of wave generators (flap- and piston-type, respectively), the experimental procedures are similar. The wave generator is programmed to create the temporal surface elevation as described by the NLSE solution at fixed position $x^*$ to first-order in steepness

$$\eta_{wave\ maker}(x^*, t) = \text{Re}(\psi(x^*, t)\exp[i(\omega t - kx^*)]\ ). \qquad (9)$$

Since the maximal compression occurs at *x=0*, a negative value for $x^*$ has to be chosen in order to observe the nonlinear soliton interaction and breather-type focusing process in the wave facility. The larger $|x|$, the more the two solitons move away from each other. The second-order Stokes correction is considered when comparing the collected data with the theoretical NLSE predictions at the respective gauge location $x_g^*$, that is

$$\eta(x_g^*, t) = \text{Re}\left(\psi(x_g^*, t)\exp[i(\omega t - kx_g^*)] + \frac{1}{2}k\psi^2(x_g^*, t)\exp[2i(\omega t - kx_g^*)]\ \right). \quad (10)$$

Note that when programming the wave maker to produce the surface elevation to first-order in steepness, results are expected to be identical as the bound waves (higher-order Stokes harmonics) are immediately generated within half a wavelength due to the intrinsic feature of the nonlinearity in the description of water waves. Moreover, fixing two key physical parameters, namely wave amplitude *a* and the carrier frequency $f = \frac{\omega}{2\pi}$ are sufficient to determine all physical features of the surface elevation. The choice for the specific values of the carrier amplitude and frequency is restricted to the stroke and frequency range specifications of the wave generator. The wave steepness *ka*, which is an indicator for the nonlinearity of the carrier wave, can be easily determined using the dispersion relation (2). One crucial step consists in scaling the solution $\Psi(X, T)$ to a dimensional form $\psi(x, t)$ satisfying Eq. (1). Considering a scaling with respect to the space- or time-NLSE does not have a major impact on the evolution of the degenerate solution in a water wave flume [37].

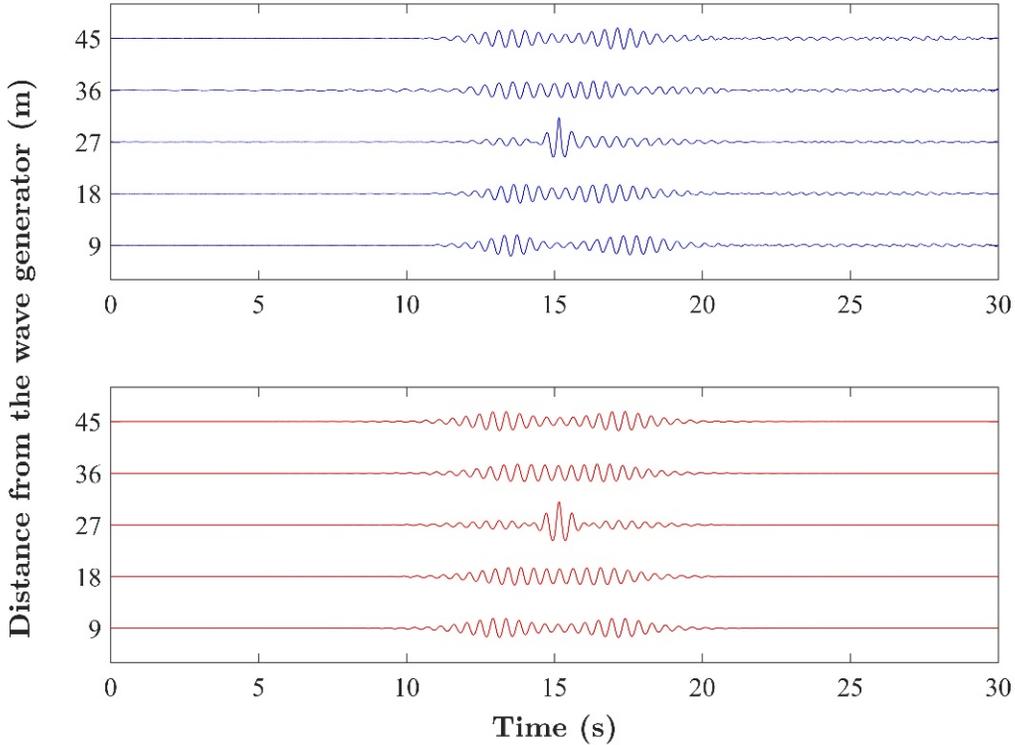

**Figure 3: Experimental observation of a degenerate soliton for *a*=0.006 m, *ka*=0.12 and *x**=-27 as measured in the Hamburg University of Technology flume.** Top: Water surface as measured by the wave gauges. Bottom: NLSE predictions at the same physical locations using Eq. (10).

The first experiment reported here aims to demonstrate the evolution of the solution over a significantly large distance of 45 m in order to observe the nonlinear and solution-specific interaction between the two envelope solitons yielding an extreme localization. On the other hand, the evolution in the Hamburg University of Technology flume was restricted to 15 m (when taking out the beach installation, effectively 12 m). To overcome this limitation, the reflection-free wave measurement at 9 m was re-injected to the wave generator four times mimicking continuation of the wave propagation. The results of these tests are shown in Fig. 3.

These results are a clear confirmation of degenerate soliton dynamics on the water surface. Note the excellent agreement in the distinct dynamics with the theoretical prediction, especially considering the total evolution distance of about 144 times the value of the wavelength.

There are obvious limitations of the NLSE model for water waves [38-40]. In fact, when waves become steep, the spectral broadening applies the natural restriction to the NLSE reducing its ability to accurately describe the wave hydrodynamics. However, this strongly depends on the initial carrier steepness and the bandwidth of the wave train [41].

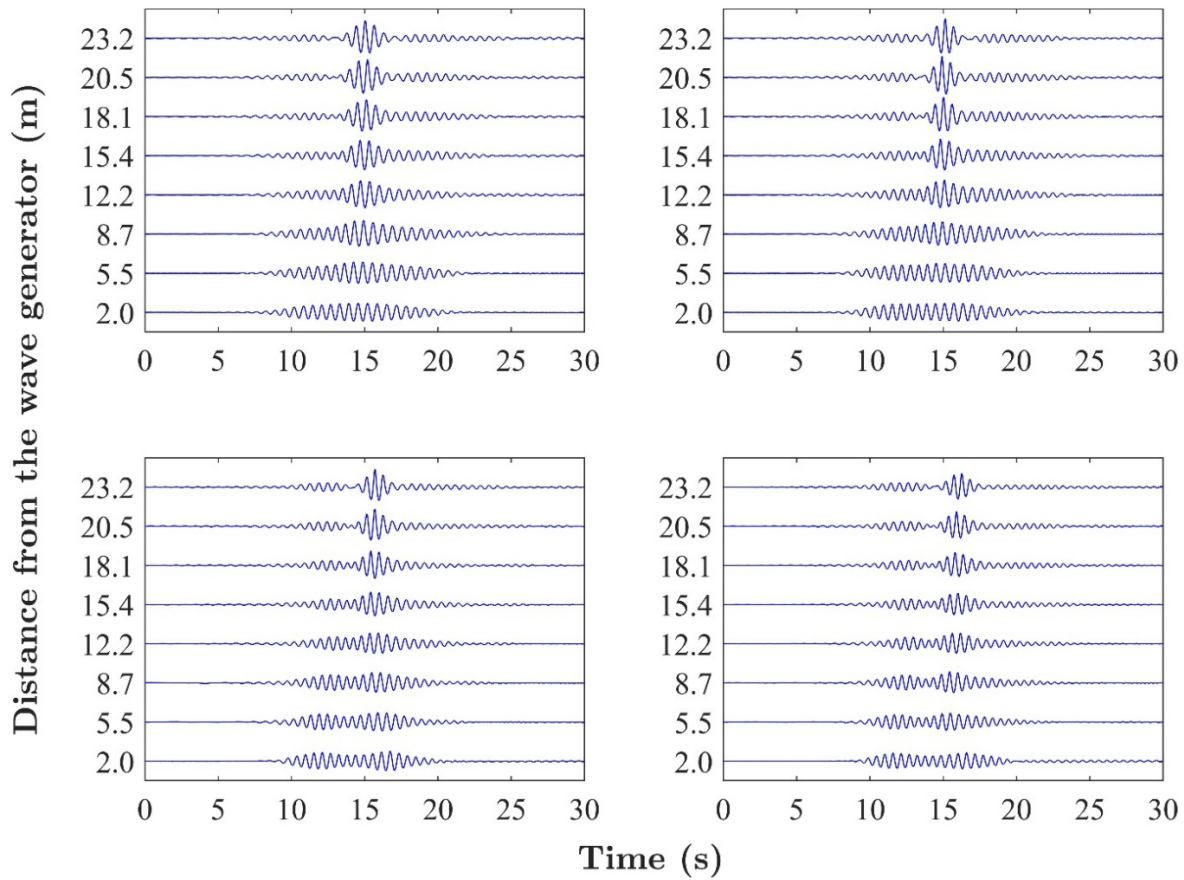

**Figure 4: Experimental observation of a degenerate soliton for *a*=0.01 m and $x^*$=-23 with varying steepness values as measured in the University of Sydney flume.** Top left: *ka*=0.10. Top right: *ka*=0.11. Bottom right: *ka*=0.12. Bottom left: *ka*=0.13.

The next series of tests have been conducted at the University of Sydney wave flume. These addressed the role of wave steepnes on the collision process. Several tests have been conducted by gradually increasing the wave steepness from 0.10 to 0.13 with a 0.01 step for the same carrier amplitude of 0.01 m. The four examples of evolution of the degenerate solution at different steepness values are shown in Fig. 4.

Here, we can clearly see that the increase of carrier steepness distorts the *clean* and ideal evolution of the NLSE solution, particularly when the carrier steepness values exceed 0.13. Consequently, the soliton interaction becomes asymmetric with a distortion of the envelope shape at the peak. These restrictions can be accurately addressed and quantified numerically by solving the Euler equations as is discussed in the next section.

## 4  Numerical simulations

The numerical simulation is performed within the framework of the potential Euler equations using the High-Order Spectral Method (HOSM) following [42]. The HOSM simulations include $2^{10}$ grid points in the physical space and a twice larger number in the Fourier domain. The iterations in time are performed with the help of a split-step Fourier procedure. The order of nonlinearity is set to $M = 6$. This corresponds to the solution that is accurate of up to 7-wave nonlinear interactions. The initial-value problem is solved in a periodic spatial domain. The wave steepness is the only physical parameter which controls the wave evolution. The steepness is determined by the quantity $ka$, where $a$ is the amplitude of the envelope solitons long before they start to collide.

With the purpose of comparing the results of the simulations with the NLSE solution (7), the computed surface evolution was transformed to the co-moving dimensionless variables (4) as used in the NLSE. It is then re-scaled to provide the unit amplitudes of the envelope solitons when these are detached at $T \to -\infty$, according to the transformations similar to (4)

$$X = 2ak(x - c_g t), \quad T = a^2 \frac{\omega}{2} t, \quad \Psi = \frac{k}{\sqrt{2}} \frac{\eta}{a}. \tag{11}$$

Note that in (11) the function $\Psi(X,T)$ is now real-valued. Three cases of the wave steepness were simulated, which correspond to $ka = 0.05$, $ka = 0.10$ and $ka = 0.15$. In all these cases, the initial condition is specified according to the solution (8). The dimensionless time is chosen to be $T_0 \approx -12$. This choice corresponds to the situation when solitons already exhibit partial overlap as can be seen in Fig. 5. This overlap seeds the interaction process in the simulations.

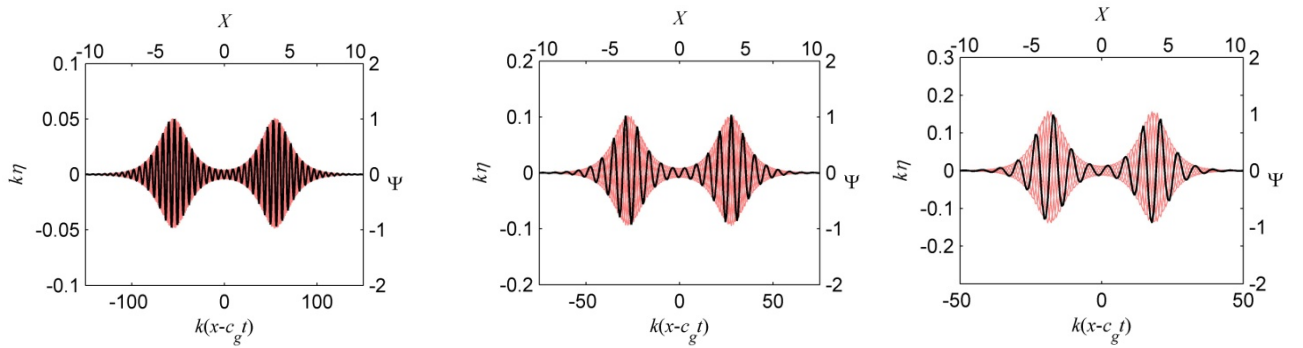

**Figure. 5. Initial conditions for the numerical simulations of the degenerate soliton.** Left: $ka = 0.05$. Middle: $ka = 0.1$. Right: $ka = 0.15$. The axes show the physical scaled coordinate and surface displacement (*y*-axis left), the standard NSLE coordinate $X = 2ak(x - c_g t)$ (*x*-axis) and the complex amplitude $\Psi = \frac{k}{\sqrt{2}} \frac{\eta}{a}$ (*y*-axis right).

The physical time of the start of the simulation $t_0$ depends on the wave steepness, see Eq. (11). In fact, it corresponds to about 340 wave periods in the steepest case shown in the right panel of Fig. 5, and to about 3000 periods in the small-amplitude case shown in the left panel of Fig. 5. In order to initiate the simulation of the HOSM code, the surface displacement and the surface velocity potential are calculated from $\psi(x, -t_0)$ with a more precise definition than Eq. (10), using the third-order asymptotic solution for nonlinear modulated waves, see [43]. Only the cases without wave breaking were simulated, thus, no filters are required to take into account wave breaking effects.

Six runs of the numerical simulations were performed with different complex phases of the initial condition $\psi(x, -t_0)$, for several wave steepness conditions. The envelope $\Psi_{env}(X, T)$ is calculated as the maximal values of $\Psi$ among these six simulations at every $X$ and $T$. The surface displacements of the initial conditions are plotted in Figs. 5 with respect to two versions of the dimensionless space and amplitude variables.

A false color representation of the evolution of each degenerate soliton envelope in time and space is shown in Fig. 6.

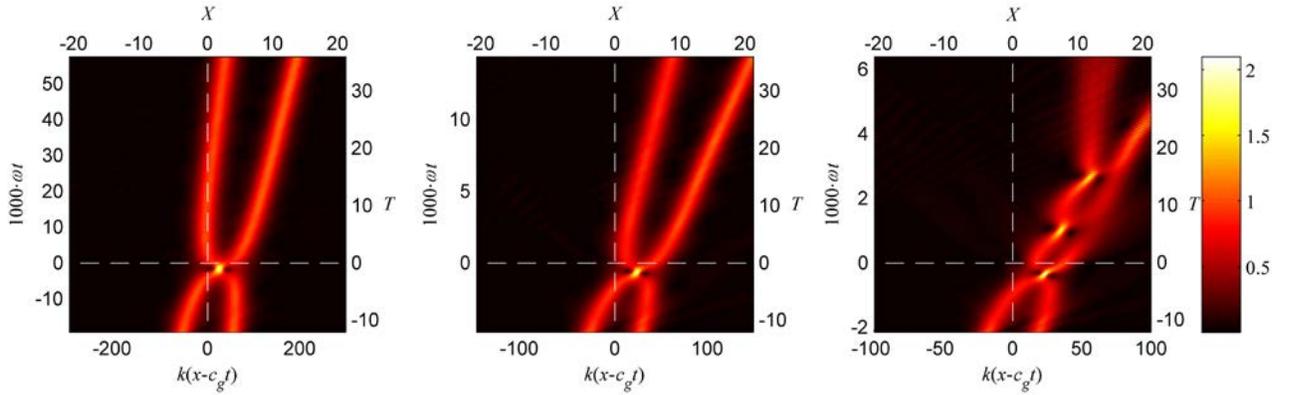

**Figure. 6. Numerical simulations of the Euler equations of the degenerate soliton for different steepness for different initial conditions.** Left: $ka$ = 0.05. Middle: $ka$ = 0.1. Right: $ka$ = 0.15. The color-coded evolution of the wave envelope $\Psi_{env}$ is shown.

The intersection of the white dashed lines corresponds to the point in time and space where the maximum wave is expected within the NLSE framework. Qualitatively, the evolution of waves with small steepness $ka$ = 0.05 (see Fig. 6 left panel) is similar to the one obtained from the NLSE (Fig. 1 right panel) and in the laboratory experiment (Fig. 3). In simulations, the two solitary groups separated initially collide, form an extreme event and then, separate again restoring their soliton shape. However, the strongly nonlinear simulation results in faster propagation of the wave groups and slightly quicker formation of the large wave (yellow dot). Interestingly, the amplitudes of the solitons after the separation are slightly different: the amplitude of the leading group is larger. The described features of the strongly nonlinear simulation becoming more pronounced when the steepness is larger than $ka$ = 0.1 (Fig. 6, middle panel).

Indeed, when the steepness further increases, $ka$ = 0.15, the new recurrence effects are becoming more apparent (Fig. 6 right panel). Moreover, when the two soliton groups merge, they form a bound state similar to the bi-soliton described in [1,32]. However, in contrast to the bi-soliton, the interaction here is asymmetric. The two subsequent extreme events are still large in amplitude in this type of recurrent dynamics. Fig. 7 shows the time evolution of the maxima of the wave elevation for the three simulations shown in Fig.6.

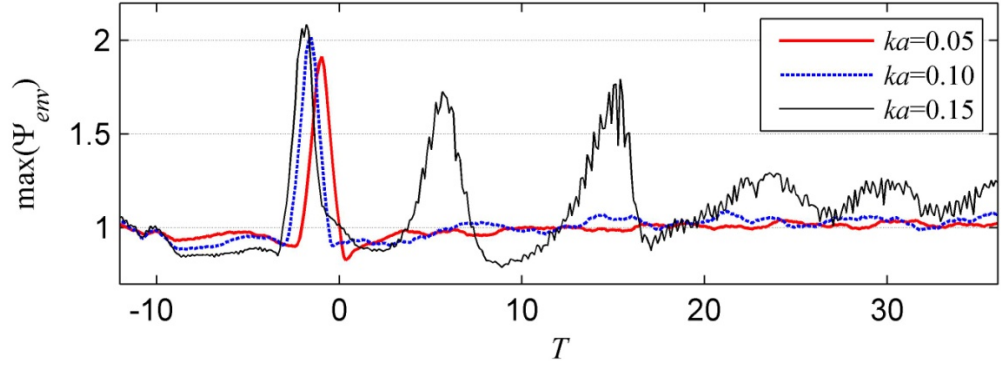

**Figure. 7. Evolution of the wave maxima in the numerical simulations shown in Fig. 6.**

After a few beating cycles, the solitary groups finally decouple. At the end of the interaction process the leading soliton has a higher in amplitude than the trailing one. After the three collisions, the envelope solitons are completely separated. The groups emerged after the third collision are not stationary. The leading soliton reveals the breathing dynamics (this can be seen in Fig. 7 for $T>20$). The second solitary group spreads decaying in amplitude. Thus, the water wave dynamics of very steep degenerate solitons shows the survival of only one (leading) soliton. Its amplitude increases while the energy of the other group reduces.

The extreme wave groups with highest amplitude which arise in the course of the wave dynamics are shown in Fig. 8.

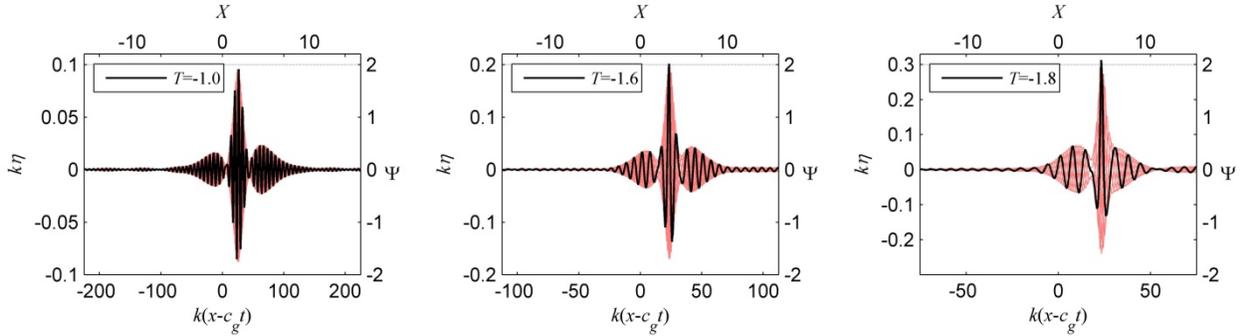

**Figure. 8. Extreme events with highest amplitude during the evolution presented in Fig. 6.**

In contrast to the envelopes shown in Fig. 5, these wave profiles possess strong back-to-front asymmetry. Some alteration of the central feature with maximal amplitude, when the wave steepness grows, may be noticed as well. This difference from the experimental observations is most-probably caused by dissipative effects [20]. The maximum water elevation is slightly smaller than anticipated by the NLSE solution for $\Psi = 2$ (marked by the dotted lines in Fig. 8) in the smaller wave steepness case, shown in Fig. 8 (left panel). However, this limit is slightly exceeded in the case of larger wave steepness shown in Fig. 8 right panel. The wave groups in Fig. 8 possess noticeable vertical asymmetry in the steeper cases (Fig. 8 middle and right panel) due to the bound (phase-locked) waves. While the wave crest exceeds the value of $\Psi = 2$ in the steepest wave case (Fig. 8 right panel), the deepest wave trough is well under the level of the NLSE solution. We emphasize that the wavelength of the carrier wave is assumed to be sufficiently large so that the capillary effects may be neglected while being small enough to satisfy the deep-water condition.

# 5 Conclusions

We have reported for the first time the experimental observation of the *degenerate* soliton interaction in nonlinear physics. This coherent structure can be considered to be a PB on the zero-background limit. The experimental data and numerical simulations are both in excellent agreement for small and moderate steepness values. This fact confirms the accuracy of the NLSE in the description of extreme wave events. Future studies will be devoted to higher-order soliton degeneracy beyond the collision of two solitons. This will improve our understanding of such effects in formation of rogue waves.

# 6 Conflict of Interest

The authors declare that the research was conducted in the absence of any commercial or financial relationships that could be construed as a potential conflict of interest.

# 7 Author Contributions

A.C. conducted the experiments in the water wave flumes. A.S. performed the HOSM simulations. All Authors designed the experiments, analysed the data and interpreted results, and took part in writing the manuscript.

# 8 Funding

A.S. was partially supported by Laboratory of Dynamical Systems and Applications NRU HSE, of the Ministry of Science and Higher Education of the Russian Federation Grant 075-15-2019-1931; and by the Grant 18-02-00042 of the Russian Foundation for Basic Research. B.K. and J.M.D. acknowledge support from the French National Research Agency (EUR EIPHI ANR-17-EURE-0002 and PIA2/ ISITE-BFC, ANR-15-IDEX-03, "Breathing Light" and "Nextlight" projects).

# 9 Acknowledgments

A.C. acknowledges Zachary Benitez and Theo Gresley-Daines for technical support. A.S. acknowledges support from the International Visitor Program of The University of Sydney.